\documentclass[letterpaper, 10 pt, conference]{ieeeconf}  

\IEEEoverridecommandlockouts                              

\usepackage{graphicx} 
\usepackage{cite}
\usepackage{amsmath,amssymb,amsfonts}
\usepackage{algorithmic}
\usepackage{algorithm}
\usepackage{graphicx}
\usepackage{textcomp}
\usepackage{xcolor}
\usepackage{bm} 
\usepackage[flushleft]{threeparttable}

\newcommand\norm[1]{\Vert#1\Vert}

\newtheorem{definition}{\textbf{Definition}}[section]

\title{\LARGE \bf
A Practitioner's Guide to Automatic Kernel Search for Gaussian Processes in Battery Applications
}

\author{Huang Zhang$^{1,2}$, Xixi Liu$^{2}$, Faisal Altaf$^{1}$ and Torsten Wik$^{2}$
\thanks{$^{1}$Department of Electromobility, Volvo Group Trucks Technology, 405 08 Gothenburg, Sweden
        {\tt\small huang.zhang@volvo.com, faisal.altaf@volvo.com}}%
\thanks{$^{2}$Department of Electrical Engineering, Chalmers University of Technology, 412 96 Gothenburg, Sweden
        {\tt\small huangz@chalmers.se, xixil@chalmers.se, torsten.wik@chalmers.se}}%
}

\begin{document}
\bstctlcite{IEEEexample:BSTcontrol} 

\maketitle
\thispagestyle{empty}
\pagestyle{empty}

\begin{abstract}
Gaussian process (GP) models have been used in a wide range of battery applications, in which different kernels were manually selected with considerable expertise. 
However, to capture complex relationships in the ever-growing amount of real-world data, selecting a suitable kernel for the GP model in battery applications is increasingly challenging.
In this work, we first review existing GP kernels used in battery applications and then extend an automatic kernel search method with a new base kernel and model selection criteria.
The GP models with composite kernels outperform the baseline kernel in two numerical examples of battery applications, i.e., battery capacity estimation and residual load prediction. Particularly, the results indicate that the Bayesian Information Criterion may be the best model selection criterion as it achieves a good trade-off between kernel performance and computational complexity.
This work should, therefore, be of value to practitioners wishing to automate their kernel search process in battery applications.
\end{abstract}

\section{Introduction}
Lithium-ion batteries have been widely adopted as energy storage systems in applications of electric vehicles and electrical grids due to their outstanding characteristics, such as high energy density and efficiency, possibilities of different power-to-energy ratios as well as decreasing costs~\cite{schmuch2018performance}. With the ever-increasing availability of different kinds of data in these battery applications, substantial efforts have been made in data-driven methods to predict battery state of health, renewable energy production, and load demand subjected to uncertainties over the last years~\cite{rauf2022machine}~\cite{wazirali2023state}. In particular, Bayesian methods, such as Gaussian processes (GPs), offer a principled approach to handling these uncertainties~\cite{rasmussen2006gaussian}. Specifically, Bayesian approaches incorporate estimates of uncertainty into the prediction with a confidence interval that consists of probabilistic upper and lower bounds. The resulting confidence intervals can be essential for decision-making under uncertainties. 

Through extracting specific input features, there has been an increasing amount of literature on developing GP regression models in battery applications, which can be divided into two categories. One is battery health estimation and remaining useful life (RUL) prediction using lab data~\cite{richardson2017gaussian}~\cite{richardson2019battery}~\cite{liu2019modified}~\cite{liu2021future} or field data~\cite{aitio2021predicting}, and the other is power prediction in microgrids with battery storage~\cite{gan2020data}~\cite{najibi2021enhanced}~\cite{yoo2024modeling}.

For battery health estimation and RUL prediction, Richardson et al.~\cite{richardson2017gaussian}~\cite{richardson2019battery} proposed GP regression models for randomized load profiles, in which 10 different composite kernels were created as sums and products of 4 base kernels (squared exponential, Mat\'ern $3/2$, Mat\'ern $5/2$, and Periodic). It was found that composite kernels based on Mat\'ern kernels provided the best prediction performance. Liu et al.~\cite{liu2019modified} also proposed GP regression models for battery health prediction for various operating temperatures and depth-of-discharge conditions, in which one composite kernel was created as the product of 3 base kernels (squared exponential, polynomial, and Laplacian) with the Arrhenius law embedded. It was found that the modified composite kernel considering knowledge of the battery mechanism outperformed the base kernel, squared exponential. In a subsequent paper by them~\cite{liu2021future}, a migrated mean function was designed and incorporated into the GP regression model to predict battery health considering knee occurrence. The prediction performance of migrated GP regression models with 3 different base kernels (squared exponential, Mat\'ern $5/2$, and rational quadratic) were compared and it was found that the Mat\'ern $5/2$ kernel provided the best performance.

For power prediction in microgrids, Gan et al.~\cite{gan2020data} used a GP regression model with a base kernel for solar photovoltaic (PV) production and load demand prediction in interconnected microgrids. The resulting prediction performance was found to improve by sharing information among microgrids. Najibi et al.~\cite{najibi2021enhanced} also used GP regression models with the Mat\'ern $5/2$ kernel for PV production prediction in 5 different PV power plants. By extracting the best features from meteorological data as inputs, GP regression models outperformed other state-of-the-art prediction methods. Considering the significant impact of residual load forecast errors on microgrid operation, Yoo et al.~\cite{yoo2024modeling} proposed GP regression models with the automatic relevance determination kernel for estimating residual load forecast errors. The GP regression model outperformed the copula-based counterpart. The aforementioned works demonstrate the applicability of GP regression models in battery applications and also the importance of the kernel. 
However, in all the above studies, different kernels have been manually selected with considerable expertise, which is becoming more and more challenging with the ever-growing amount of real-world data.

In this work, we first review existing GP kernels that have been used in battery applications from the perspective of a practitioner. Then for each application category, i.e., battery health estimation and RUL prediction in electric vehicles, and power prediction in microgrids, we conduct comparative studies of automatic kernel search using three different model selection criteria through two numerical examples. Specifically, our \textbf{key results and contributions} are summarized as follows:
\begin{itemize}
\item 
The GP regression models with composite kernels found using the automatic kernel search method outperform the baseline kernel (Mat\'ern $5/2$) in two numerical examples of battery applications with respect to high accuracy and reliable uncertainty quantification.
\item
Among three model selection criteria for GPs, the Bayesian Information Criterion (BIC) may be the best to find the best composite kernel as it achieves a good trade-off between kernel performance and computational complexity in the two numerical examples. The Laplace approximation is comparable in quality but compromises for computational speed and potential inconsistencies.
\end{itemize} 


\section{Gaussian Process Regression}
The goal of a supervised learning problem is to learn input-output mappings from a training set $\mathcal{D}$ of $n$ observations, i.e., $\mathcal{D} = \{(\bm{x}_{i}, y_{i})|i=1, \ldots, n\}$, where $\bm{x}_i \in \mathbb{R}^{p}$ denotes an input vector of dimension $p$ and $y_i \in \mathbb{R}$ denotes a scalar output. The output can either be continuous, as in the regression case, or discrete as in the classification case~\cite{bishop2006pattern}. In this practitioner's guide, we are only concerned with Gaussian process models for regression problems.  

To simplify modeling situations, we take the underlying model in the form of $y=f(\bm{x}) + \varepsilon$, where $f(\bm{x})$ denotes a latent deterministic function and $\varepsilon \sim \mathcal{N}(0, \sigma^2_{\varepsilon})$ is additive independent identically distributed Gaussian noise. 

From the function-space view~\cite{rasmussen2006gaussian}, the function $f(\bm{x})$ is also a random variable that follows a particular distribution. Here, we assume that the function $f(\bm{x})$ is distributed as a Gaussian process (GP), i.e.,
\begin{equation}
    f(\bm{x}) \sim \mathcal{GP}(m(\bm{x}), \kappa(\bm{x}, \bm{x}')),
\end{equation}
where input vectors $\bm{x}$ and $\bm{x}'$ are both in either the training set or the test set. $m(\bm{x})$ and $\kappa(\bm{x}, \bm{x}')$ are the mean and covariance functions, respectively, defined as
\begin{align}
m(\bm{x}) &= \mathbb{E}[f(\bm{x})]\\
\kappa(\bm{x}, \bm{x}') &= \mathbb{E}[(f(\bm{x})-m(\bm{x}))(f(\bm{x}')-m(\bm{x}'))].
\end{align}
For simplicity, the prior mean function $m(\bm{x})$ is often assumed to be zero. The covariance function $\kappa(\bm{x}, \bm{x}')$ is also called the kernel function, which defines the covariance between any two function values. In this case, the GP is solely determined by $\kappa(\bm{x}, \bm{x}')$ that is parameterized by hyperparameters $\bm{\theta}$.

For all the training input points $\bm{X}=[\bm{x}_1, \bm{x}_2, \ldots, \bm{x}_n]$, the GP defines a prior probability distribution that is jointly Gaussian, i.e.,
\begin{equation}\label{Gaussian_prior}
   \mathbf{f}  \sim \mathcal{N}(\bm{0}, \bm{K}(\bm{X}, \bm{X})),
\end{equation}
where $\mathbf{f}=[f(\bm{x}_1), f(\bm{x}_2), \ldots, f(\bm{x}_n)]^T$, $\bm{K}(\bm{X}, \bm{X})$ denotes the $n \times n$ covariance matrix evaluated at all pairs of training points. The prior distribution of noisy outputs $\bm{y}$ can be expressed by
\begin{equation}\label{eq5}
    \bm{y} \sim \mathcal{N}(\bm{0}, \bm{K}(\bm{X}, \bm{X})+\sigma_{\varepsilon}^2 \bm{I}),
\end{equation}
where $\bm{y} = [y_1, y_2, \ldots, y_n]^T$ denotes $n$ observed training outputs, $\bm{I}$ denotes the identity matrix of size $n$, and $\sigma_{\varepsilon}^2$ is the noise variance. 

For $m$ test input points $\bm{X}^*=[\bm{x}^*_1, \bm{x}^*_2, \ldots, \bm{x}^*_m$], the joint prior distribution of the observed training outputs and the function values at the test points can be expressed by~\cite{rasmussen2006gaussian}
\begin{equation} \label{eq6}
\begin{bmatrix}
\bm{y} \\
\mathbf{f}^*
\end{bmatrix} \sim 
\mathcal{N}
(\bm{0}, 
\begin{bmatrix}
\bm{K}(\bm{X}, \bm{X}) + \sigma_{\varepsilon}^2 \bm{I} & \bm{K}(\bm{X}, \bm{X}^*)\\
\bm{K}(\bm{X}^*, \bm{X}) & \bm{K}(\bm{X}^*, \bm{X}^*)
\end{bmatrix}),
\end{equation}
where $\mathbf{f}^* = [f(\bm{x}^*_1), f(\bm{x}^*_2), \ldots, f(\bm{x}^*_{m})]^T$, $\bm{K}(\bm{X}, \bm{X}^*)$ denotes the $n \times m$ covariance matrix evaluated at all pairs of training and test points, and similarly for $\bm{K}(\bm{X}^*, \bm{X})$ and $\bm{K}(\bm{X}^*, \bm{X}^*)$.
 
\subsection{Hyperparameter Optimization} 
The covariance function $\kappa(\bm{x}, \bm{x}')$ can be optimized on the training data by maximizing the log marginal likelihood (LML) defined as~\cite{rasmussen2006gaussian}
\begin{equation}\label{eq_LML}
    \mathcal{L} = \log p(\bm{y} | \bm{X}),
\end{equation}
where $p(\bm{y} | \bm{X})$ is the marginal likelihood (or model evidence) and is the integral of the likelihood times the prior.
Under the Gaussian prior defined in Eqn. (\ref{Gaussian_prior}), the likelihood is a factorized Gaussian. So the integral becomes analytically tractable, which yields
\begin{equation}
    \mathcal{L} = \underbrace{-\frac{1}{2} \bm{y}^T \bm{K}_{\bm{y}}^{-1} \bm{y}}_\text{data fit} - \underbrace{\frac{1}{2} \log|\bm{K}_{\bm{y}}|}_\text{complexity penalty}
                     - \underbrace{\frac{n}{2} \log 2\pi}_\text{normalization constant},
\end{equation}
where $\bm{K}_{\bm{y}}=\bm{K}+\sigma_{\varepsilon}^2 \bm{I}$ is the covariance matrix for the noisy outputs $\bm{y}$ and $\bm{K} = \bm{K}(\bm{X}, \bm{X})$ is the covariance matrix for the noise-free function $\mathbf{f}$. Although the LML is commonly used for hyperparameter optimization, it does not explicitly penalize superfluous hyperparameters if used as a model selection criterion, thus leading to potential overfitting. We will discuss this problem further in the following section.

\subsection{Inference}

With possibly optimized hyperparameters in the covariance function, the predictive distribution (or GP posterior) can be calculated by conditioning the joint Gaussian prior distribution (Eqn. (\ref{eq6})) on the observations using the conditional distributions of the multivariate normal distribution (see Theorem proof in Ref.~\cite{statproofbook}) as
\begin{equation}
    \mathbf{f}^*|\bm{X}, \bm{y}, \bm{X}^*  \sim \mathcal{N}(\overline{\mathbf{f}}^*, \mathrm{cov}(\mathbf{f}^*))
\end{equation}
with
\begin{align}
    \overline{\mathbf{f}}^* = & \bm{K}(\bm{X}^*, \bm{X}) \bm{K}_{\bm{y}}^{-1} \bm{y}\\
    \mathrm{cov}(\mathbf{f}^*) = & \bm{K}(\bm{X}^*, \bm{X}^*) - \bm{K}(\bm{X}^*, \bm{X}) \bm{K}_{\bm{y}}^{-1} \bm{K}(\bm{X}, \bm{X}^*)  
\end{align}
where $\overline{\mathbf{f}}^*$ denotes the corresponding mean values and $\mathrm{cov}(\mathbf{f}^*)$ denotes the covariance matrix. If the prior mean function is non-zero, the posterior mean becomes
\begin{equation}
    \overline{\mathbf{f}}^* = \bm{m}(\bm{X}^*) + \bm{K}(\bm{X}^*, \bm{X}) \bm{K}_{\bm{y}}^{-1} (\bm{y}-\bm{m}(\bm{X})).
\end{equation}
To compute the predictive distribution for noisy test outputs $\bm{y}^*$, simply add the noise variance $\sigma_{\varepsilon}^2 \bm{I}$ to $\mathrm{cov}(\mathbf{f}^*)$.

\section{Automatic Kernel Search Method}
Gaussian process models use a covariance function (also called kernel function) to define the covariance between any two function values
\begin{equation}
    \mathrm{cov}(f(\bm{x}), f(\bm{x}')) = \kappa(\bm{x}, \bm{x}').
\end{equation}
The kernel function specifies the similarity between function values at two inputs $\bm{x}$ and $\bm{x}'$. The prior on the noisy observations is expressed by Eqn. (\ref{eq5}).

\subsection{Base Kernels and Operations}\label{SubsecA}
In this subsection, we will introduce some commonly-used kernels in battery applications as base kernels, which can then be combined to express different priors over $f$. If we let $r = \norm{\bm{x} - \bm{x}'}$, then these base kernels are expressed by 
\begin{itemize}
\item 
Squared exponential (SE) 
\begin{equation}
    \kappa_{\mathrm{SE}}(r) = \exp \left( -\frac{r^2}{2 \ell^2} \right),
\end{equation}
\item 
Mat\'ern $5/2$ (Ma5)
\begin{equation}
    \kappa_{\mathrm{Ma5}}(r) = \left( 1+\frac{\sqrt{5}r}{\ell}+\frac{5r^2}{3 \ell^2} \right) \exp \left( - \frac{\sqrt{5}r}{\ell} \right)
\end{equation}
\item 
Periodic (Pe)
\begin{equation}
    \kappa_{\mathrm{Pe}}(r) = \exp \left( -2 \frac{\sin^2(\pi r/p)}{\ell^2} \right)
\end{equation}
\item 
Linear (Lin)
\begin{equation}
    \kappa_{\mathrm{Lin}}(r) = \bm{x} \cdot \bm{x}'
\end{equation}
\item 
Rational quadratic (RQ)
\begin{equation}
    \kappa_{\mathrm{RQ}}(r) = \left( 1 + \frac{r^2}{2\alpha \ell^2}\right)^{-\alpha}
\end{equation}
\end{itemize}
where $p$ denotes the period length, $\ell$ denotes the characteristic length-scale, and $\alpha$ is the shape parameter determining length-scales' diffuseness. The SE kernel is infinitely differential and is thus very smooth. However, such strong assumptions about the smoothness of the function do not hold when modeling many physical processes in reality. Therefore, one may resort to the Ma5 kernel that is twice differentiable in the mean-square sense. The Pe kernel allows modeling periodic functions with the period length $p$. The Lin kernel computes the inner product in input space. The RQ kernel can be seen as a scale mixture (or an infinite sum) of SE kernels with different characteristic length-scales.
\begin{definition}[Stationary Kernel Functions]\label{def_stationary}
    A stationary kernel function is a function that is invariant to translations in input space, i.e., its values depend only on the difference $\bm{x} - \bm{x}'$.
\end{definition}

\begin{definition}[Non-Stationary Kernel Functions]\label{def_non_stationary}
    A non-stationary kernel function is a function that is variant to translations in input space, i.e., its values are different whenever input vectors are different.
\end{definition}

According to Definitions \ref{def_stationary} and \ref{def_non_stationary}, the SE, Ma5, Pe, and RQ kernels are stationary while the Lin kernel is non-stationary. Here, a set of base kernels $\mathcal{K}=\{ \text{SE}, \text{Ma5}, \text{Pe}, \text{Lin}, \text{RQ} \}$ is considered. Note that for clarity reasons, the scaling of each kernel $\sigma_f$ is omitted. 



The sum or the product of two valid kernels (i.e., positive semidefinite kernels) is still a valid kernel, which allows a wide range of kernels to be constructed via additions ('$+$') and multiplication ('$\times$') of commonly-used base kernels. In one-dimensional cases, sums of different base kernels can model the superposition of multiple processes at different scales, while products of different base kernels can transform a global structure into a local one. In multi-dimensional cases, sums of base kernels can model additive structures over different dimensions, while products of base kernels can model smooth structures. Here, a set of operations $\mathcal{O}=\{ +, \times \}$ is considered.

\subsection{Model Selection Criteria}
The model selection for GPs includes choices of base kernels and operations, and settings of kernel hyperparameters. A general rule for model selection is preferring simpler models over competing ones that explain the data equally well, often referred to as Occam's razor~\cite{rasmussen2000occam}. A model selection criterion should consider this rule to achieve a good trade-off between model performance and complexity.

Simplistically, the optimized LML expressed by
\begin{equation}
    \hat{\mathcal{L}}=\log p(\bm{y} | \bm{X}, \hat{\bm{\theta}})
\end{equation} 
can be used to evaluate the model quality of GPs with optimal hyperparameters $\hat{\bm{\theta}}$. However, if it is used as a model selection criterion, more complex models that allow overfitting will be favored because it does not explicitly penalize the number of hyperparameters. To avoid this, one could integrate the likelihood over all the hyperparameters $\bm{\theta}$ to obtain the model evidence:
\begin{equation}\label{model_evidence}
    \mathcal{Z} = p(\bm{y} | \bm{X}) = \int p(\bm{y} | \bm{X},\bm{\theta}) p(\bm{\theta}) d \bm{\theta}.    
\end{equation}
The integral above may not be analytically tractable and in general one may resort to approximation techniques, such as Akaike Information Criterion (AIC)~\cite{akaike1974new} and Bayesian Information Criterion (BIC)~\cite{schwarz1978estimating}. The AIC approximated log model evidence is defined as~\cite{bishop2006pattern}
\begin{equation}
    \log\mathcal{Z}_\text{AIC} = \hat{\mathcal{L}} - m,
\end{equation}
where $m$ is the number of hyperparameters, while the BIC approximated log model evidence is defined as~\cite{bishop2006pattern}
\begin{equation}
    \log\mathcal{Z}_\text{BIC} = \hat{\mathcal{L}} - \frac{m}{2} \log n,
\end{equation}
where $n$ is the number of observations in the training set $\mathcal{D}$. To compensate for the overfitting of more complex models, the AIC explicitly penalizes the number of hyperparameters, and this penalty term is added to the LML. In contrast, the BIC introduces a larger penalty depending on both the number of hyperparameters and the number of observations. However, neither AIC nor BIC considers the uncertainty in the hyperparameters, and therefore their approximations are rather crude. To address this, the Laplace approximation aims to find a Gaussian approximation of the marginal likelihood (\ref{model_evidence}) using a second-order Taylor approximation around its optimum. The Laplace approximated log model evidence is defined as~\cite{bishop2006pattern}
\begin{equation}
    \log\mathcal{Z}_\text{Lap} = \hat{\mathcal{L}} + \log p(\hat{\bm{\theta}}) + \frac{m}{2} \log(2\pi) - \frac{1}{2} \log(|\bm{H}|),
\end{equation}
where $\bm{H}=-\nabla \nabla \log(p(\bm{y} | \bm{X},\bm{\theta}) p(\bm{\theta})) |_{\bm{\theta}=\hat{\bm{\theta}}}$ is the Hessian matrix evaluated at $\hat{\bm{\theta}}$~\cite{bishop2006pattern}, which can be computed using automatic differentiation in most machine learning libraries~\cite{baydin2018automatic}.
Here, a set of approximation techniques (or model selection criteria) $\mathcal{S}=\{\log\mathcal{Z}_\text{AIC}, \log\mathcal{Z}_\text{BIC}, \log\mathcal{Z}_\text{Lap} \}$ is considered.

\subsection{Kernel Search Algorithms}
The successful deployment of GP models in battery applications greatly depends on the selected kernel, which requires considerable expertise. However, in the era of big data, the manual selection of an appropriate kernel for a GP model has become exceedingly challenging for users with limited expertise as the underlying structures within real-world datasets typically exhibit complexity beyond what commonly-used base kernels can capture, such as the ones introduced in Section \ref{SubsecA}. Therefore, to address this issue, automatic kernel search algorithms have been proposed to find the best composite kernel from the training data in an iterative and greedy process, and each possible composite kernel was scored by the BIC after first optimizing the LML $\hat{\mathcal{L}}$, such as Compositional Kernel Search (CKS)~\cite{duvenaud2013structure} and Automatic
Bayesian Covariance Discovery (ABCD)~\cite{lloyd2014automatic}. These algorithms have then been extended to become scalable to big data, such as Scalable Kernel Composition (SKC)~\cite{kim2018scaling} and Concatenated Composite Covariance Search (3CS)~\cite{berns20213cs}.

In battery applications, evaluating all possible sums and products of base kernels easily becomes computationally formidable. Therefore, instead of the combinatorial search over all possible kernel combinations, we use a greedy search described in Ref.~\cite{duvenaud2013structure} and summarized in Algorithm \ref{alg}. 
Specifically, at the first level, the base kernel in the set $\mathcal{K}$ (see Section \ref{SubsecA}) with the highest approximated model evidence value on the training data ($\text{ApproxModelEvidence}(\mathcal{D}, \kappa)$ in Algorithm \ref{alg}) is selected as the best one $\kappa^*$ with optimal hyperparameters $\hat{\bm{\theta}}$. At the next level, we first create composite kernels ($\kappa_n$) from a set of base kernels $\mathcal{K}$, a set of operations $\mathcal{O}$, and the best kernel $\kappa^*$ with optimal hyperparameters $\hat{\bm{\theta}}$ from the previous level ($\text{CreateKernel}(\kappa^*, \kappa, o)$ in Algorithm \ref{alg}), and then we select the composite kernel with the highest approximated model evidence value at this level. This searching process continues until the level reaches its maximum level of search $L$. Finally, the optimal composite kernel is returned together with its approximated model evidence value. Considering the small-sized training data ($n<10K$) in the following two numerical examples, and our experience with the GP regression models, setting the maximum level of search to 3 should be sufficient to achieve a satisfactory trade-off between model performance and complexity.
Notably, optimizing over hyperparameters of composite kernels is not a convex problem. To alleviate the issue of local optima, the hyperparameters that belong to the best kernel from the previous level are initialized to their previously optimized values and the newly introduced hyperparameters are initialized with zeros.

\begin{algorithm}[htpb]
\caption{Greedy Search for Optimum Composite Kernel~\cite{duvenaud2013structure}}
\label{alg}
\begin{algorithmic}[1]
\renewcommand{\algorithmicrequire}{\textbf{Input:}}
\renewcommand{\algorithmicensure}{\textbf{Output:}}
\REQUIRE A set of base kernels $\mathcal{K}$, a set of operations $\mathcal{O}$, a model selection criteria $s \in \mathcal{S}$, a training set $\mathcal{D}$, and maximum level of search $L$
\ENSURE Composite kernel $\kappa^*$ with the lowest value $s^*$ 
\\ \textit{Initialization} : $\kappa^* = \emptyset$, $s^* = - \infty$\\
\STATE Level $n=1$
\FOR {each kernel $\kappa \in \mathcal{K}$}
\STATE $s = \text{ApproxModelEvidence}(\mathcal{D}, \kappa)$
\IF {$s > s^*$}
\STATE $\kappa^* = \kappa$
\STATE $s^* = s$
\ENDIF
\ENDFOR

\FOR {each level $n = 2$ \textbf{to} $L$}
\FOR {each kernel $\kappa \in \mathcal{K}$}
\FOR{each operation $o \in \mathcal{O}$}
\STATE $\kappa_n = \text{CreateKernel}(\kappa^*, \kappa, o)$
\STATE $s = \text{ApproxModelEvidence}(\mathcal{D}, \kappa_n)$
\IF {$s > s^*$}
\STATE $\kappa^*_+ = \kappa_n$
\STATE $s^* = s$
\ENDIF
\ENDFOR
\ENDFOR
\STATE $\kappa^* = \kappa^*_+$
\ENDFOR
\RETURN $\kappa^*$, $s^*$
\end{algorithmic}
\end{algorithm}

\section{Numerical Examples}
We compare the different composite kernels found using three model selection criteria and benchmark them to the state-of-the-art kernel in two numerical examples. In both examples, we name the best composite kernel found using AIC, BIC, and Laplace approximation to be CK-AIC, CK-BIC, and CK-Lap, respectively. The arrows behind performance evaluation metrics denote if lower ($\downarrow$) or higher ($\uparrow$) values are better, and the best of them are bold.
\subsection{Example 1 - Battery Capacity Estimation}
\subsubsection{Battery aging dataset}
\begin{figure}[thpb]
  \centering
  \includegraphics[width=0.9\linewidth]{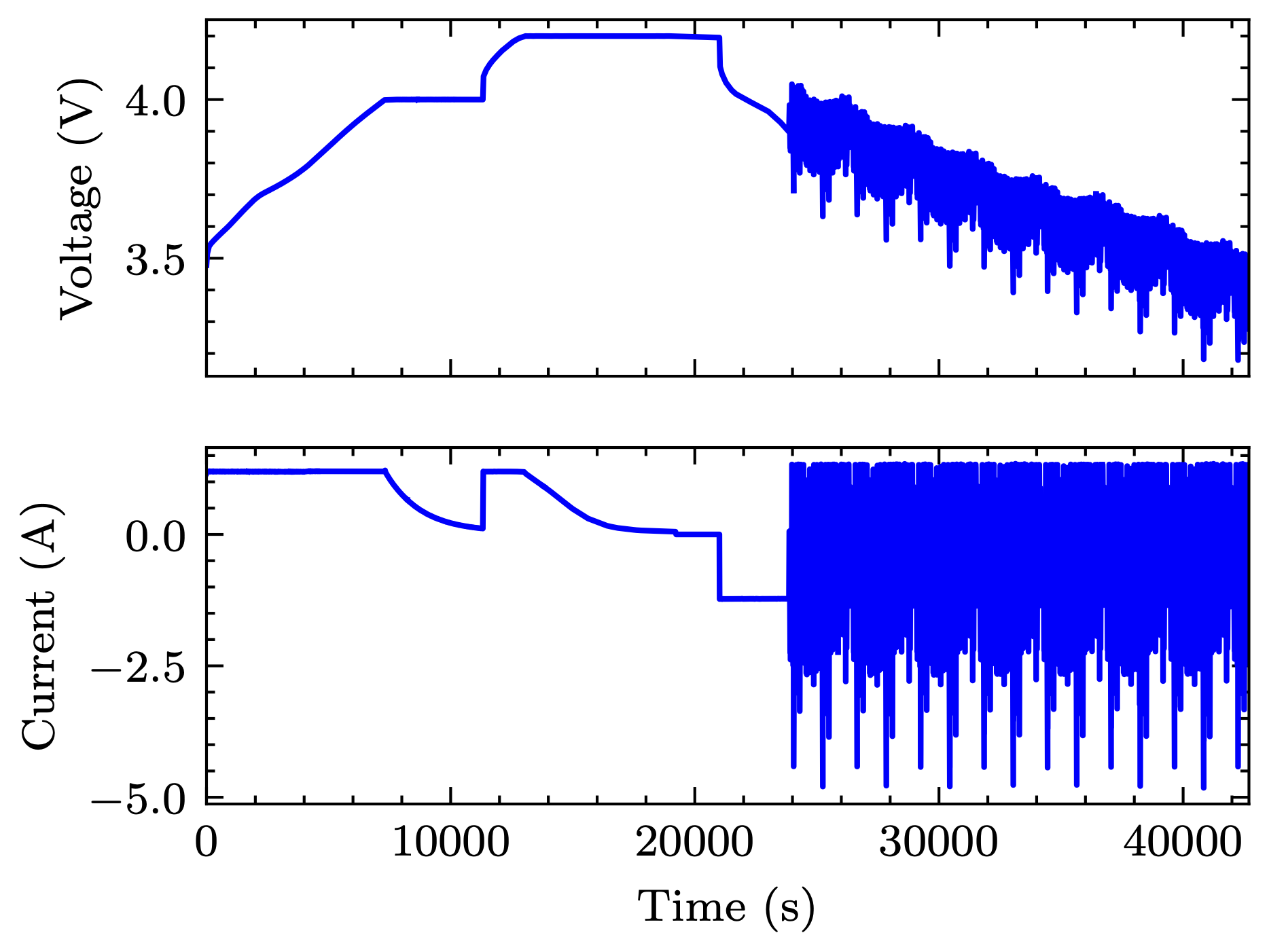}
  \caption{One full charge-discharge cycle of a sample cell [W4] in the dataset.}
  \label{fig3}
\end{figure}

To demonstrate the effectiveness of the automatic kernel search method in the battery capacity estimation problem, an open-source battery dataset generated by Stanford Energy Control Laboratory is used here~\cite{pozzato2022lithium}. In total, this dataset comprises 10 lithium nickel manganese cobalt oxide (NMC)/graphite-silicon cylindrical cells manufactured by LG Chem (model INR21700-M50T, 4.85 Ah nominal capacity). The test purpose is to characterize battery aging behaviors under electric vehicle real-driving profiles. All the cells were first charged with one of 4 different constant current (CC) C-rates (C/4, C/2, 1C, and 3C) until the voltage reached 4V, and then constant voltage (CV) discharged until the current reached the cutoff value of 50 mA. Next, cells were identically CC-CV charged at C/4 until the voltage reached 4.2V, i.e., 100\% state-of-charge (SoC). Subsequently, cells were identically discharged at C/4 from 100\% to 80\% SoC, and then discharged with the Urban Dynamometer Driving Schedule (UDDS) driving profile to 20\% SoC (see Fig.~\ref{fig3}). All the cells were cycled at a constant ambient temperature of 23$^{\circ}$C. Time-series cell voltage and current were continuously measured, and two battery health metrics, i.e., rated capacity (C/20 discharge, 23$^{\circ}$C), and internal resistance (from Hybrid Pulse Power Characterization tests) were measured every 25 or 50 cycles.

\subsubsection{Feature engineering}
To develop GP regression models for the battery capacity estimation, we select the Oxford 3-feature set proposed by Greenbank et al.~\cite{greenbank2021automated}. Specifically, this feature set consists of 3 features extracted from full cycling data, i.e., time spent between voltages corresponding to 1st and 33rd percentiles over every 20 hours ($V_{12}$), time spent between voltages corresponding to 33rd and 67th percentiles over every 20 hours ($V_{23}$), and calendar time ($t$). The output variable is capacity change ($\Delta Q$) over every 20 hours.

\subsubsection{Train-test split}
\begin{figure}[thpb]
  \centering
  \includegraphics[width=0.9\linewidth]{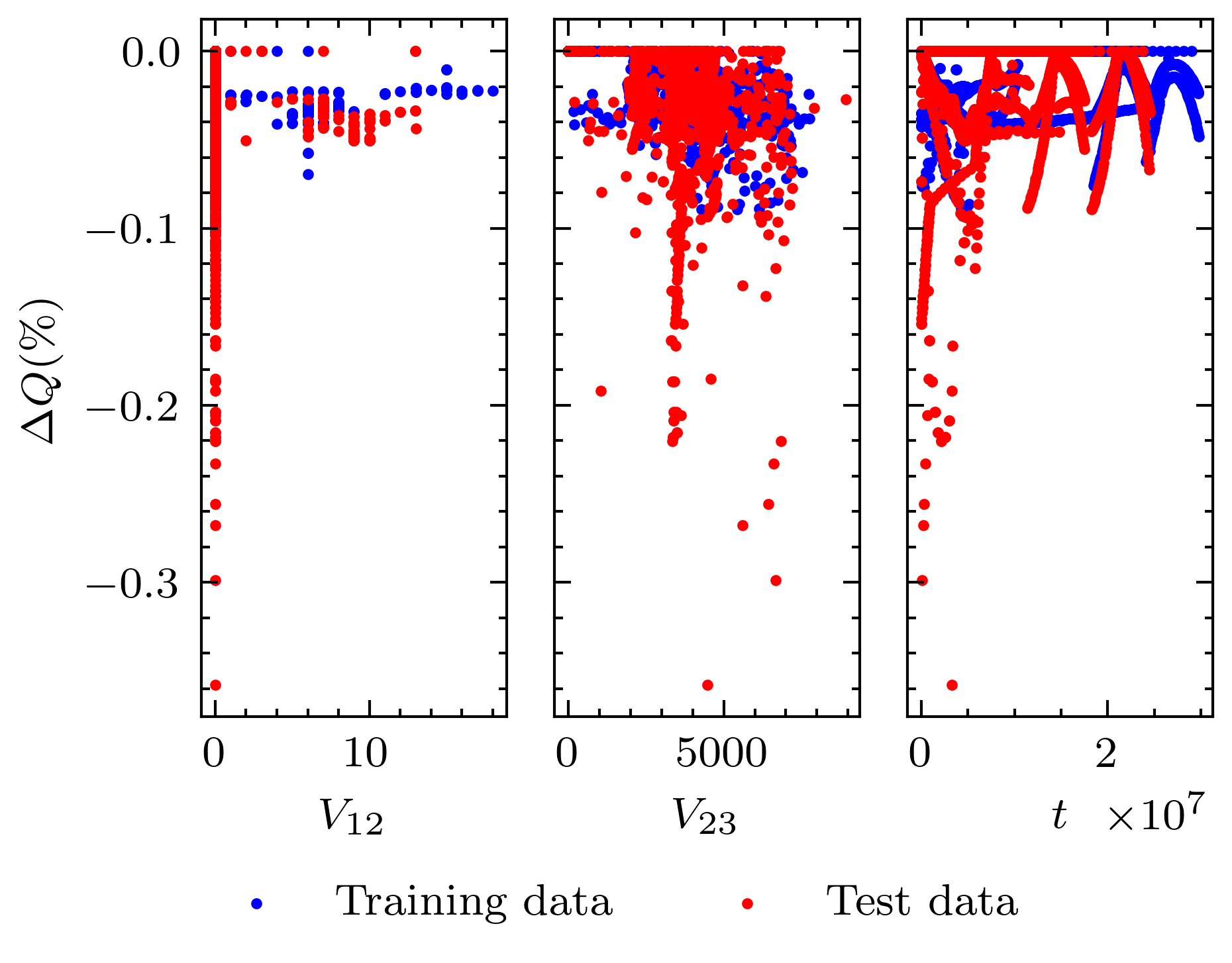}
  \caption{Capacity change versus Oxford 3 features at one stratified train-test split.}
  \label{fig3.5}
\end{figure}

To improve model generalization performance and guarantee reliable model evaluation on the test set, the stratified random sampling method~\cite{reitermanova2010data} is used for the train-test split. In the dataset, there are 4 different charge C-rates, i.e., C/4, C/2, 1C, and 3C. Therefore, the charge C-rate is used as the criterion to first classify cells into fast-charged (1C and 3C) cells, and normal-charged (C/4, C/2) cells. Then equal ratios of fast-charged and normal-charged cells are kept in the training set (4 cells) and test set (4 cells). To illustrate the effectiveness of the feature engineering method, 3 features versus the output variable at one stratified train-test split are plotted in Fig.~\ref{fig3.5}. It can be seen that there is a small amount of test data that dooes not overlap with the training data. Note that the train-test split is repeated 5 times with their results averaged to reduce the randomness of the outcome.

\subsubsection{Kernel search and performance evaluation}
\begin{figure}[thpb]
  \centering
  \includegraphics[width=0.9\linewidth]{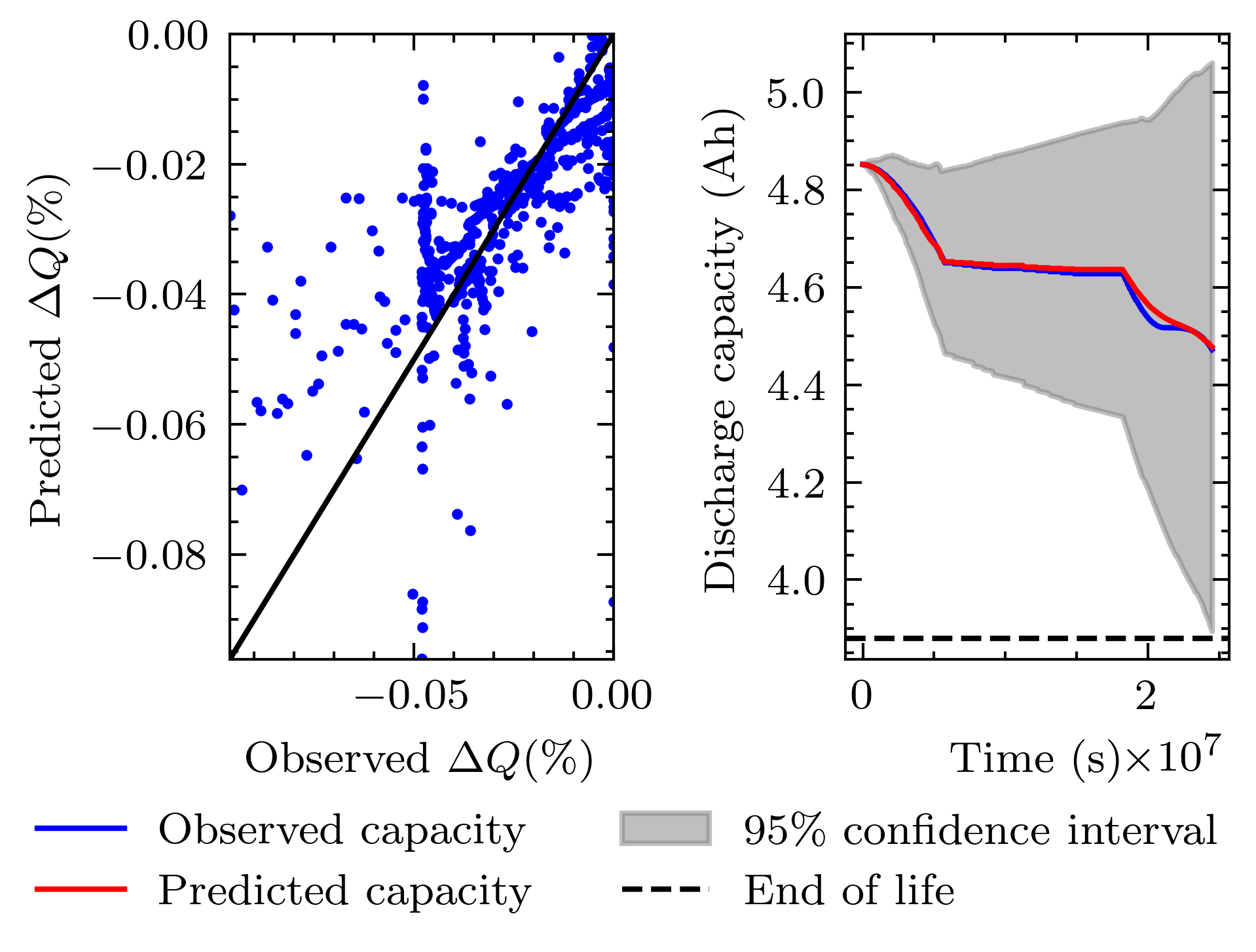}
  \caption{Predicted versus observed $\Delta Q$ (left) and predicted capacity versus time (right) of a sample cell [W8] in the test set. Note that the composite kernel (Ma5 $\times$ Ma5 $\times$ Pe) found using Laplace approximation is used here.}
  \label{fig4}
\end{figure}

In this first numerical example, the Mat\'ern $5/2$ is selected as the benchmark kernel as it has shown excellent performance in battery health estimation and remaining useful life (RUL) prediction problem~\cite{richardson2019battery}~\cite{liu2021future}.
The results of GP regression models with the Mat\'ern $5/2$ kernel and three composite kernels over 5 train-test splits are summarized in Table~\ref{tab1}. 
In terms of battery capacity point prediction performance measured by root-mean-square error (RMSE) and mean absolute percentage error (MAPE), it can be seen that the composite kernel found using Laplace approximation (CK-Lap) performs the best over CK-AIC, CK-BIC, and the Mat\'ern $5/2$ kernel. In terms of battery capacity range prediction performance measured by mean prediction interval width (MPIW) and prediction interval coverage probability (PICP), it can be seen that the Mat\'ern $5/2$ kernel performs the best over the others as its PICP value is closest to nominal coverage probability (95.4\%) and MPIW value is the smallest among all. The CK-Lap kernel is overconfident with its PICP value larger than 95.4\% but is still closer to it than CK-AIC or CK-BIC kernel.
In addition to predicted capacity fade curves with high accuracy and reliable uncertainty quantification, knee-onset and knee points on the capacity fade curve are also captured in the composite kernel (see Fig.~\ref{fig4}).

In this example, the goal was to develop a battery capacity estimation model with high accuracy, reliable uncertainty quantification, and consideration of possible knee occurrence on the capacity fade curve. Although the test data does not strongly overlap with the training data as illustrated in Fig.~\ref{fig3.5}, the GP regression models with composite kernels extrapolate capacity changes well outside the training data range. Furthermore, the experimental results in Table~\ref{tab1} suggest that the composite kernel found using Laplace approximation as the model selection criterion is the best candidate to achieve this goal, even though its range prediction is a bit overconfident for the investigated realistic electric vehicle driving profile (i.e., UDDS).

\begin{table}[ht]
\caption{GP regression model performance for capacity estimation}
\label{tab1}
\begin{center}
\resizebox{\columnwidth}{!}{
\begin{threeparttable}
\begin{tabular}{|l|c|c|c|c|}
\hline
\textbf{Kernel} & \multicolumn{2}{|c|}{\textbf{Point prediction}} & \multicolumn{2}{|c|}{\textbf{Range prediction}} \\
\cline{2-5} 
& RMSE (\%) $\downarrow$ & MAPE (\%) $\downarrow$ & PICP (\%) $\uparrow$ & MPIW (\%) $\downarrow$\\
\hline
Ma5 & 0.12  & 2.17 & \textbf{95.43} & \textbf{0.41} \\
\hline
CK-AIC\tnote{1} & 0.11 & 2.08 & 90.99 & 0.42 \\
\hline
CK-BIC\tnote{2} & 0.11 & 2.08 & 90.99 & 0.42 \\
\hline
CK-Lap\tnote{3} & \textbf{0.08} & \textbf{1.49} & 96.07 & 0.42 \\
\hline
\end{tabular}
\begin{tablenotes} 
    \item[1 2] The best composite kernel over 5 train-test splits is found to be RQ $+$ Pe $+$ RQ using AIC and BIC.  
    \item[3] The best composite kernel over 5 train-test splits is found to be Ma5 $\times$ Ma5 $\times$ Pe using Laplace approximation.
\end{tablenotes}
\end{threeparttable}
}
\end{center}
\end{table}

\subsection{Example 2 - Residual Load Demand Prediction}
\subsubsection{Grid-connected photovoltaic battery system dataset}
To demonstrate the effectiveness of the automatic kernel search method in the residual load demand prediction problem (load demand - renewable energy production), a residential grid-connected photovoltaic (PV) battery system for a housing association of 132 households in Gothenburg, Sweden, is studied here. The residential PV battery system comprises a stationary battery energy storage system (BESS) that contains 14 lithium-ion battery packs retired from electric buses and a PV generation unit. The specification of the PV battery system is summarized in Table~\ref{tab2}.

\begin{table}[ht]
\caption{PV battery system parameters}
\label{tab2}
\begin{center}
\begin{tabular}{|l|c|c|}
\hline
\textbf{Parameter} & \textbf{Unit} & \textbf{Value}\\
\hline
PV peak power ($P^{\mathrm{max}}_{\mathrm{PV}}$) & kWp & 170.8\\
Grid power limit ($P^{\mathrm{lim}}_{\mathrm{grid}}$) & kW & 100\\
Battery rated capacity ($E_b$) & kWh & 200\\
Battery maximum charge/discharge power ($P^{\mathrm{lim}}_b$) & kW & 70\\
Battery round-trip efficiency ($\eta$) & \% & 96\\
SoC window ($\mathrm{SoC}_{\mathrm{max}}-\mathrm{SoC}_{\mathrm{min}}$) & \% & 85-20 \\
Calendar life ($L_{\mathrm{cal}}$) & years & 13.5 \\
Cycle life ($L_{\mathrm{cyc}}$) & EFC & 6000\\
\hline
\end{tabular}
\end{center}
\end{table}

\subsubsection{Feature engineering}
To develop GP regression models for residual load demand prediction, we must consider the specific usage of GP regression models in grid-connected PV battery systems. Considering that the electricity spot prices for the next 36 hours are released at 13:00 every day on the Nordic market, it would be beneficial if the residual load demand for the next 36 hours is also predicted at 13:00 every day so that the optimal control policy can be computed. 
Therefore, the corresponding input features and output variables in vector forms are constructed as $[P_\mathrm{res}(t-23), \ldots, P_\mathrm{res}(t), W(t), D(t), H(t)]$ and $[P_\mathrm{res}(t+1)]$, respectively. Here, $W(t)$ denotes the week in a year at time $t$, $D(t)$ denotes the day in a week at time $t$, and $H(t)$ denotes the hour in a day at time $t$.

\subsubsection{Train-test split}
The 3-month data in 2022 (2022-10-01 - 2022-12-31) is used as the training set, and the 3-month data in 2023 (2023-10-01 - 2023-12-31) is used as the test set.
\subsubsection{Kernel search and performance evaluation}
In this second numerical example, Mat\'ern $5/2$ is again selected as the benchmark kernel as it has shown excellent performance in PV production prediction problem~\cite{najibi2021enhanced}.
The results of GP regression models with the Mat\'ern $5/2$ kernel and three composite kernels are summarized in Table~\ref{tab3}. 
Interestingly, the kernel search processes using Laplace approximation and BIC as the model selection criteria were terminated early before the maximum level of search ($L=3$), since the approximated log model evidence of all composite kernels at the next level is less than that of the best kernel found at the previous level. In terms of residual load demand point prediction performance measured by RMSE, it can be seen that CK-Lap performs better than CK-AIC, CK-BIC, and the baseline Mat\'ern $5/2$ kernel. In terms of residual load demand range prediction performance measured by MPIW, it can be seen that CK-BIC performs better than the others. However, all these kernels are underconfident with their PICP values less than nominal coverage probability (95.4\%).

In this example, the goal is to develop a residual load prediction model with high accuracy and reliable uncertainty quantification. In particular, large prediction errors have been shown to lead to a lower operation economy and accelerated battery aging in microgrids~\cite{chen2019impacts}. In this regard, the experimental results in Table~\ref{tab3} indicate that all three composite kernels found using the automatic kernel search method are better choices to achieve this goal than the baseline kernel for the stochastic load prediction in microgrids here.
\begin{table}[ht]
\caption{GP regression model performance for load prediction}
\label{tab3}
\begin{center}
\resizebox{\columnwidth}{!}{
\begin{threeparttable}
\begin{tabular}{|l|c|c|c|c|}
\hline
\textbf{Kernel} & \multicolumn{2}{|c|}{\textbf{Point prediction evaluation}} & \multicolumn{2}{|c|}{\textbf{Range prediction evaluation}} \\
\cline{2-5} 
& RMSE (kW) $\downarrow$ & MAPE (\%) $\downarrow$ & PICP (\%) $\uparrow$ & MPIW (kW) $\downarrow$\\
\hline
Ma5 & 14.63  & 35.95 & 76.56 & 37.44 \\
\hline
CK-AIC\tnote{1} & 13.19  & 26.49 & 78.70 & 33.63  \\
\hline
CK-BIC\tnote{2} & 12.83  & \textbf{25.51} & 79.60 & \textbf{33.17}  \\
\hline
CK-Lap\tnote{3} & \textbf{12.49}  & 27.60 & \textbf{82.00} & 33.41  \\
\hline
\end{tabular}
\begin{tablenotes} 
    \item[1] The composite kernel is found to be (Ma5 $+$ Ma5) $\times$ Ma5 using AIC.
    \item[2] The composite kernel is found to be Ma5 $+$ Ma5 using BIC.
    \item[3] The kernel is found to be RQ using Laplace approximation.
\end{tablenotes}
\end{threeparttable}
}
\end{center}
\end{table}

\begin{figure}[thpb]
  \centering
  \includegraphics[width=0.9\linewidth]{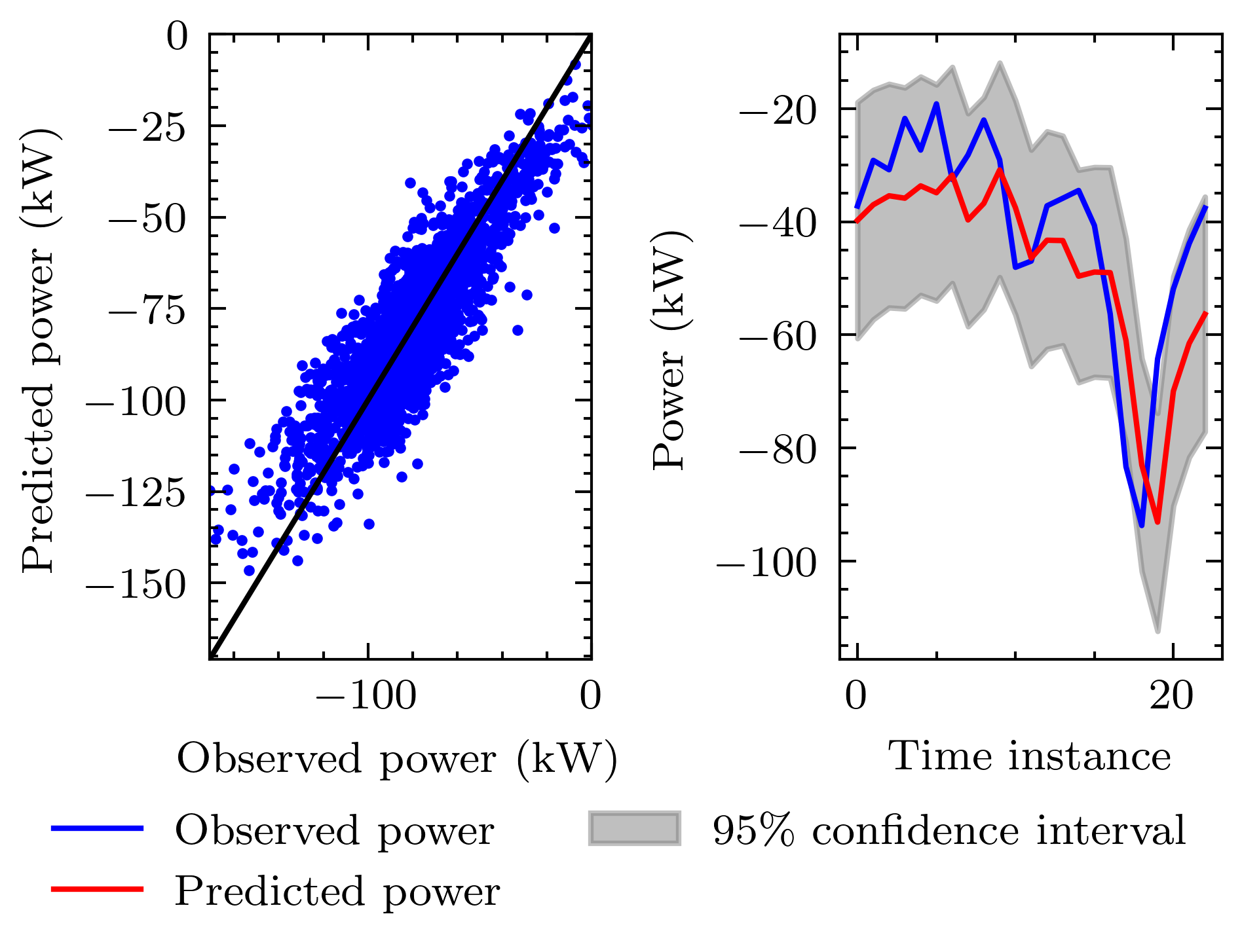}
  \caption{Predicted versus observed residual load demand (left) and predicted residual load demand versus time (right) over 24 hours in the test set. Note that the RQ kernel found using Laplace approximation is used here.}
  \label{fig5}
\end{figure}

\section{Conclusion}
Gaussian process (GP) regression models have been used for a wide range of battery applications, for example, battery health estimation and remaining useful life (RUL) prediction, renewable energy production, and load demand prediction. Different GP kernels have been manually selected for these problems, which requires considerable expertise. To capture complex relationships in real-world data, we resort to an existing automatic kernel search method to find the best composite kernel for GPs in battery applications. In particular, three model selection criteria for GPs, i.e., Akaike Information Criterion (AIC), Bayesian Information Criterion (BIC), and Laplace approximation, were compared using this automatic kernel search method.
With the aid of the automatic kernel search method, it has been demonstrated that the GP regression models with the composite kernels performed better than the baseline kernel (Mat\'ern $5/2$) in both numerical examples of battery applications. Specifically, the GP regression model with the composite kernels found using all three model selection criteria can provide outstanding battery health estimation and RUL prediction performance with high accuracy, reliable uncertainty quantification, and excellent shape approximation of capacity fade curves, while the GP regression model with the composite kernels using all three model selection criteria can also provide better residual load demand prediction than the baseline kernel in terms of accuracy and uncertainty quantification. Among the three model selection criteria for GPs, the BIC may be the best as it achieves a good trade-off between model performance and computational complexity. In contrast, the Laplace approximation is comparable to the BIC. However, it is much more expensive due to the Hessian computation, which will become computationally formidable when the automatic kernel search method is extended to big data.

These findings provide the following insights for our future research, i.e., 1) to further improve the prediction performance of GP regression models, migration concepts based on sharing information among batteries with similar aging behaviors, or microgrids within the same region; 2) to make the automatic kernel search method to become scalable to big data in battery applications, requiring different Hessian approximations to be investigated.






\section*{ACKNOWLEDGMENT}
This work was supported by the Swedish Energy Agency (Grant number P2024-00998).


\bibliographystyle{IEEEtran}
\bibliography{References}

\end{document}